\documentclass[12pt]{iopart}
\usepackage{graphicx}

\begin{document}
\title {The Fractal Dimension of Ionization Cascades in the Glow Discharge }
\author {Reginald D. Smith \\ Bouchet-Franklin Research Institute \\ P.O. Box 360262 \\ Decatur, GA 30036-0262}
\date {October 18, 2004}
\ead {reggiesmith@alumni.virginia.edu}

\begin {abstract}

 The glow discharge's main ionization breakdown processes have been
understood for about one hundred years. The glow discharge, however, 
still remains as an avenue of fresh research in pattern formation 
and far from equilibrium systems. The primary and secondary 
ionization processes can be mathematically modeled as general 
branching processes. Not only is the Townsend breakdown criterion 
obtained but the ionization avalanche can be characterized as a 
branching set with a unique Hausdorff fractal dimension. These 
fractal dimensions show applicability using similarity principles 
and Paschen's Law.

\end {abstract}
\maketitle

The cylindrical geometry, planar electrode DC glow discharge is one of the most familiar features in the study of ionized gases. 
Though the features of the DC discharge have been described since at least the time of Michael Faraday,
they were not theoretically understood until the work of John  S.E. Townsend early in the 20th century \cite{Howatson, Raizer}.  Townsend was able to describe the gas breakdown in terms to two ionization cascades. A primary ionization cascade of electrons in the gas as voltage is increased due to electron avalanches beginning with electrons from atoms ionized by external radiation like UV sources, natural radioactivity, or cosmic rays. A secondary ionization cascade is caused by electron ionizations from the positive ion counterparts to the electrons colliding with other atoms in the cathode or other gas atoms. At breakdown, the discharge becomes self-sustaining, transitioning from the dark Townsend discharge to the luminous glow discharge. At this point secondary ionization provides enough initial electrons and external radiation is no longer needed.
This has been long understood, however, the glow discharge has recently become a fertile area 
of research into pattern formation and non-equilibrium systems \cite {selforg1, selforg2,selforg3}. 
In particular, \v{S}ija\v{c}i\'{c} and Ebert \cite{glowdis, glowdis2} have formulated the Townsend breakdown in terms of a supercritical and subcritical 
bifurcation. This paper will investigate the ionization cascades in terms of the 
topology of the ionization cascade. Specifically, by formulating the ionization avalanche as 
a branching process, it can be described with a Hausdorff fractal dimension. 

\section {General Branching Process}

The theory of branching processes is a theory of the growth of a population through 
generations given each member of the population leaves the population, remains single, or 
produces a number of progeny given the transition probabilities of a fixed Markov process. 
It was first described by Francis Galton, H.W. Watson, and earlier by Bienaym\'{e}. 
Though related research in the probabilistic nature of breakdown has been
done \cite{Wijsman, Loeb}, in this paper we will use the general branching process described by Harris \cite{Harris} to derive 
the Townsend breakdown condition.

The objects considered by our branching process will be the electrons in the gas discharge
primary and secondary ionizations. The transition probability $p_k$ is the probability that an 
electron will produce k electrons through collision from one generation to the next. The probability generating 
function $f(s)$ is 
\begin{equation}
f(s) = \sum_{k=0}^{\infty} p_ks^k
\label {genfun}
\end {equation}
Where $f'(1)$ is the defined as the expectation value, $m$, which is the 
average number of new electrons each individual electron will produce 
each generation. If $m\geq1$ there is a non-zero probability of an 
infinite number of electrons in the limit of the  number of
generations goes to infinity. According to the theory of branching 
processes, as the number of generations goes to infinity, the 
electron population trends to either infinity or zero.

The general branching process is a multitype branching process 
where an attribute of the members of the population are described
by using a random finite set of variables, x, which belongs to an 
interval X, a subset of the positive real numbers. This variable can be used 
to describe among other attributes, the age, energy, or in this 
paper, the position of a particle. The point distribution $\omega$ is 
the set of all distinct points $x_1$ to $x_k$ with positive integer weights 
$n_1$ to $n_k$. 
\begin {equation}
\omega =(x_1,n_1; x_2, n_2;...x_k, n_k)
\label {omega}
\end {equation}
The integer weight $n_k$ represents the number of objects point $x_k$.
Define $\Omega_X$ as the set of all point distributions. 
Define a set function $\tilde{\omega}$ where $\tilde{\omega}(A) = \sum_{x_i\in{A}}n_i$. 
If A is null $\tilde{\omega}(A) = 0$. In addition each subset A can have a probability 
defined $P(\tilde{\omega}(A)) = r$ which is the probability that there are r total objects in the 
subset A. Finally, a function s(x) is defined
\begin {equation}
s(x)=\int_{X}s(x)d\omega(x)=n_1s(x_1)+n_2s(x_2)+...n_ks(x_k)
\label {s}
\end {equation}
Where s(x) is 1 if x is in A and 0 otherwise. 
The moment generating functional (MGF), the function that supplements the probability generating
function in point set distributions, is defined as
\begin {equation}
\Phi(s) = Ee^{-\int s dw} = \int_{\Omega_X} e^{-\int s dw} dP
\label {MGF}
\end {equation}
but also has an alternative formulation
\begin {equation}
\Phi(s) = f_0 + \sum_{n=1}^\infty\int...\int f_n(x_1,...,x_n)e^{-s(x_1)-...-s(x_n)}dx_1...dx_n
\label {MGF2}
\end {equation}
where $f_0$ is the probability of no objects and $f_n(x_1,...,x_n)dx_1...dx_n$ is the probability of
n objects in which the first is in the interval $(x_1, x_1 + dx_1)$ etc. This is the formulation we
will use in our derivation.
The purpose of defining the MGF is to enable us to use each $x_k$ 
and its corresponding weight as the number of electrons at a point 
$x_k$ on the length of the ionization region from 0 to d, 
the length of the cathode sheath. For a general branching process the corresponding 
variable to the expectation value, $m$, is the eigenvalue $\rho$ of the first moment of the 
MGF where the first moment is defined as
\begin {equation}
M_1(x, A)=E_X Z_1(A)=\frac {-\partial \Phi_x(ts_A)} {\partial t}_{t=0}
\label {firstmoment}
\end {equation}
$Z_1(A)$ is the number of objects in A in the first generation.
The first moment also has a density $m_1$ defined as
\begin {equation}
\label {densityformula}
M_1(x, A) = \int_A m_1(x,y)dy
\end {equation}

The density $m_1$ is also related to the eigenvalue, $\rho$ which plays the same role in multitype branching processes as the expectation value, $m$, and the right eigenfunction $\mu(x)$ 
and the left eigenfunction $\nu(y)$ by
\begin {equation}
m_1 (x,y) = \rho \mu(x) \nu(y)[1 + O(\Delta)] 
\label {finalBP}
\end {equation}
\subsection {Primary Ionization}
For the primary ionization in the Townsend discharge we define the 
transition probabilities of $p_0$ being the probability of recombination,
$p_1$ the probability of scattering, and $p_2$ the probability of 
ionization. An electron is assumed to only be able to ionize a
maximum of one other electron in a collision. 

Let $\lambda$ be the mean free path for electrons in the gas. Therefore $e^{-\frac{d} {\lambda}}$ be the
probability of no collision of an electron over the distance d \cite{Loeb2}. Here we consider d to be the length of the ionization region in the cathode sheath. The probability of an electron starting
at point x colliding at a distance y is $e^{-\frac {(y-x)} {\lambda}}$. We can then define the MGF for the first
generation as
\begin {equation}
\Phi_x^{(1)} (s) = (e^{-\frac{(d-x)}{\lambda}})+\sum_{n=0}^\infty p_n \int_X e^{-\frac{(y-x)}{\lambda}}e^{-ts(y)}dy
\label {FirstionBP}
\end {equation}
which is expanded as
\begin {equation}
\Phi_x^{(1)} (s) = (e^{-\frac{(d-x)}{\lambda}}) + p_0\int_0^d e^{-\frac{(y-x)}{\lambda}}e^{0}dy + p_1\int_0^d e^{-\frac{(y-x)}{\lambda}}e^{-ts_A}dy + p_2\int_0^d e^{-\frac{(y-x)}{\lambda}}e^{-2ts_A}dy
\label {FirstionBP2}
\end {equation}
$s_A$, given the equation above, is the total number of electrons in the first generation. For the primary
cascade we assume one initiating electron so $s_A = 1$. The first moment as defined in \ref{firstmoment} is
\begin {equation}
M_1 = p_1\int_0^d e^{-\frac{(y-x)}{\lambda}}dy + 2p_2\int_0^d e^{-\frac{(y-x)}{\lambda}}dy 
\label {Primaryfirstmoment}
\end {equation}
which can be simplified as
\begin {equation}
M_1 = m\int_0^de^{-\frac{(y-x)}{\lambda}}dy
\label {Primaryfirstmoment2}
\end {equation}
where $m$ if the expectation value of the probability generating function of the primary ionization
cascade.

The density is then
\begin {equation}
m_1 = me^{-\frac{(y-x)}{\lambda}}
\label {Primarydensity}
\end {equation}
Finally, given \ref{finalBP} we can see the values of the eigenfunctions must be $\mu(x) = e^{\frac{x}{\lambda}}$
and $\nu(y) = e^{-\frac{y}{\lambda}}$ so the eigenvalue, and thus expectation value, of the general branching process
is
\begin {equation}
\rho = m
\label {Expectation1}
\end {equation}

Thus the eigenvalue $\rho$ 
equals $f'(1) = m$ of the transition probabilities as expected. The expected value of the number 
of objects, $Z$, in generation n is given by $Z_n = m^n$

According to Townsend's ionization theory, the number of electrons at 
a length d created from one primary electron is $N_d = e^{ad} - 1$ or $N_d \approx e^{ad}$. 
However, since our branching process assumes a maximum of two electrons produced every
generation, the correct base for the exponential growth is 2 not $e$.
If d is again the total length of the ionization region in the cathode sheath, 

\begin {equation}
m^n = 2^{ad}
\label {equiv}
\end {equation}

\subsection {Secondary Emission}

Though secondary emission is initiated by ions, 
since there is an equality of the number of ions and primary 
electrons we can assume the primary electrons to be the causal 
objects in the secondary emission. The transition probabilities 
for secondary emission are slightly different. $P_0$ is the probability 
an ion (primary electron) collides with the cathode and is absorbed. 
$P_1$ is the probability an ion (primary electron) collides with the 
cathode, is absorbed and emits a secondary electron. $P_1$ is equal to 
$\gamma$, Townsend's secondary ionization coefficient.

Then MGF is then

\begin {equation}
\Phi_x^{(1)} = p_0\int_0^d e^{0} dy + p_1\int_0^d e^{-ts(y)} dy
\label {secondMGF}
\end {equation}

and the first moment and density are

\begin {equation}
M_1 = p_1s_A\int_0^d e^0 dy
\label {secondfirstmoment}
\end {equation}

\begin {equation}
m_1 = ms_A
\label {secondfirstdensity}
\end {equation}

However, in this case $s_A$ is not one since we are not assuming a single parent, 
rather we are assuming all the ions from the primary electrons at all points 0 to d to be initial objects in 
secondary emission so $s_A = Z_n$. 
Given \ref{finalBP} we can see that neither x nor y are in the first moment's density 
so the left and right eigenfunctions must both be equal to 1 giving us a eigenvalue $\rho_2$ of

\begin {equation}
\rho_2 = ms_A
\label {secondeigenvalue}
\end {equation}

Given we showed earlier $Z_n = 2^{ad} -1$ and $m = \gamma$ we see

\begin {equation}
\rho_2 = \gamma(2^{ad}-1)
\label {Townsend condition}
\end {equation}

Since an infinite avalanche only happens when nonzero probability if 
$\rho_2 \geq1$ we see that we have the Townsend breakdown condition under
conditions of binary division.

\section {Branching Trees}

Now that we have shown the Townsend breakdown criterion to be a result of a 
general branching process, we can extend this result to branching trees. Let a 
branching tree, $T$, be defined as a graph G where the set of vertices $V(G)$ consists of 
all objects in all generations and let the set of edges $E(G)$ be the set of directed edges 
point from an object to its immediate "descendants" in the next generation. The branching 
tree starts with a single root at generation 0 and expands with the reproduction of the 
objects in subsequent generations.

There has been much research linking probabilistic branching processes and trees \cite {Hawkes, Lyons, Liu}
They have conclusively shown that a branching process 
can be represented by a branching tree.

The branching tree has a boundary, $\partial T$, 
which is the set of all infinite rays from the root. In other words, 
the set of all  branches on the branching tree which do not include leaves. $\partial T$ forms
a space on which lie the objects in the branching process. The distance metric is defined for two points $x, y$ in $\partial T$ as $e^{-x\wedge y}$ where $x\wedge y$ is 
the index of the closest generation in which x and y have a common ancestor. An interesting result is given this definition of a metric space, the
branching tree forms a random fractal with a fractal dimension defined as
\begin {equation}
D = \log{br[T]}
\label {dimension}
\end {equation}

\begin{figure}
\includegraphics{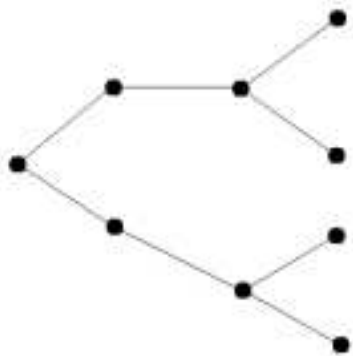}
\caption{A branching tree with four generations starting from its root.}
\label {btreepic}
\end{figure} 

$br[T]$, the branching number, is defined in general as the supremum of
a variable $\lambda$ where $\lambda^{-n}$ defines the capacity for flow
at an edge $n$ generations from the root. If $\lambda$ holds a value greater
than $br[T]$ there can be no flow in the tree. For trees generated by branching
processes, $br[T]$ is equal to the expectation value of the branching process
($\rho$). One consideration before extending the branching tree results to 
the primary and secondary ionization of the glow discharge is the 
fact that the theoretical branching tree model is a set of infinite 
rays while the breakdown processes in the glow discharge occur only 
over a finite length. It is my contention that since the ionization 
cascade is limited by an external physical constraint, the current 
limiting resistor in series with the discharge tube, that an infinite 
theoretical model can provide a close approximation for the ionization 
cascades which are supercritical as both the primary and secondary 
ionization are at breakdown.

Given the above equation for fractal dimension we can find expressions 
for the fractal dimension of both the primary and secondary ionization 
avalanches. For primary ionization we first use the expression \ref{equiv}. 
We then solve for $\log m$ to get

\begin {equation}
D_1 = \alpha \lambda \ln 2
\label {primarydimension}
\end {equation}
Where $\lambda$ is the mean free path of the gas. 
We assume the mean free path is the average distance an 
electron travels before a collision and is thus 
\begin {equation}
\lambda = d/n
\label {MFP}
\end {equation}
Where $n$ is the number of generations in the branching process. Also you can interpret $D_1$ as 
\begin {equation}
D_1 = (\lambda/\lambda_i)\ln 2
\label {primarydimension2}
\end {equation}
Where $\lambda_i$ is the average ionization length.
$D_1$ will subsequently be known as the
primary fractal dimension. 
For secondary ionization since we know $m$ is the Townsend breakdown condition, we obtain 

\begin {equation}
D_2 = \log{\gamma} + \alpha d \ln 2 
\label {secondarydimension}
\end {equation}

$D_2$ can also be expressed in terms of $D_1$

\begin {equation}
D_2 = \log{\gamma} + {nD_1} 
\label {secondarydimension2}
\end {equation}

$D_2$ will subsequently be known as the
secondary fractal dimension. Obviously $D_2$ scales with the size of the system while 
$D_1$ is dependent only on the characteristics of the gas and electric 
fields that give rise to the first ionization coefficient.

\section {Similarity Principles and Paschen's Law}

The similarity principles are a common concept in dealing with ionized gases. 
Similarity principles allow us to compare discharges of different linear dimensions. 
With similarity relations we can relate various macroscopic variables of the two 
discharges with the same potential difference drop to each other. 
The fractal dimensions of the primary and secondary ionization cascades 
both yield surprising results given the similarity principles

Assume two discharges. One has a length $d_1$ while the other has a length $d_2$ where $d_2=kd_1$. 
Assume both have the same cylindrical geometry and planar electrodes and that the voltages between the electrodes in each
are $U_1=U_2$. 
Two well known similarity principles \cite{Jones} are the similarity principle for the mean 
free path, $\lambda_1/\lambda_2 = 1/k$ and the first ionization coefficient $\alpha_1=k\alpha_2$. 
We can see that there is a corresponding similarity relation for the fractal dimension of the 
first ionization cascade

\begin {equation}
\alpha_1 \lambda_1 = \alpha_2 \lambda_2
\label {firstsimilarity}
\end {equation}

Therefore, if two discharges are similar, they have the same primary 
fractal dimension and thus share an identical underlying topology. 
Therefore, with two discharges with the same fractal dimension, all 
other similarity principles necessarily follow including $p_1d_1=p_2d_2$. 
So two discharges with identical $pd$ and $U$ also have the same primary fractal dimension. 
Thus there exists a relationship $D_1 = f(pd)$. Given

\begin {equation}
\alpha = Ape^{-Bp/E}
\label {firstionizationcoeff}
\end {equation}

And given the mean free path equation
\begin {equation}
\lambda = \frac {kT} {p \sigma}
\label {MFPequation}
\end {equation}

Where $k$ is Boltzmann's constant, $T$ is the temperature of the gas
and $\sigma$ is the cross-section.
Assuming $E = U/d$ we can deduce $f(pd)$ of being of the form 

\begin {equation}
D_1 = \frac {AkTe^{-Bpd/U}} {\sigma}
\label {similarityequation}
\end {equation}

This result explains in greater clarity the relationship between the primary fractal dimension
and the similarity principles. Given any discharge with the same gas constants, the value of
$pd/U$ is the same for all discharges with the same primary fractal dimension. Thus the similarity
principles of $p_1d_1=p_2d_2$ given $U_1=U_2$ are a subset of a larger similarity principle. 
The principle that all discharges with the same primary fractal dimension follow 

\begin {equation}
\frac {p_1d_1} {U_1} = \frac {p_2d_2} {U_2}
\label{newsimilarity}
\end {equation}

\subsection {Paschen's Law}
 
Paschen's Law is well-known as the principle that the breakdown voltage of a 
gas is constant for a constant $pd$. All points on the Paschen curve indicate 
the voltage where the gas the curve describes (assuming planar electrode geometries) 
transitions from the Townsend discharge to the normal glow. Given the results in \ref{newsimilarity} 
we can now determine that for any discharge with a given $pd$, its breakdown voltage 
occurs at the same primary fractal dimension. In addition, since $\frac {p_1d_1} {U_1} = C_1$ where C is a constant and thus
$U = C_2pd$ where $C_1=1/C_2$
we can deduce that on any Paschen curve plot by drawing a straight line of any positive slope from the
origin, all points the straight line intersects on the Paschen curve have the same primary
fractal dimension on breakdown.

The secondary fractal dimension also provides a new insight into Paschen's Law. 
By definition at every point on the Paschen curve the Townsend breakdown criterion 
is satisfied. In fact, theoretical expressions for the Paschen Curve are derived 
using the breakdown criterion as a starting point. Loeb \cite{Loeb2} in fact describes
the condition for a stable discharge according to Holst-Oosterhuis-Seeliger as

\begin{equation}
(1+\frac{1}{\gamma})=e^{\int_0^d{\alpha dx}}
\label{stability}
\end{equation}

Once the normal glow is initiated, if the gas has a value of pd greater than or 
equal to the value of $pd$ at the Paschen minimum, the gas voltage and $pd$ values 
adjust themselves to the values at the Paschen minimum. If the value of $pd$ was 
less than the minimum, the gas maintains itself at the higher breakdown value of 
voltage and the lower $pd$ value. These conditions only change once voltage and 
current return to positive differential conductivity in the abnormal glow. Thus, 
given that the normal glow always maintains itself on the Paschen curve under 
typical conditions the normal glow can be described in all states by the secondary 
ionization fractal dimension $D_2=0$. This reinterprets Paschen's Law as a contour 
line representing $D_2=0$. This is significant in that it shows the normal and abnormal (due to the stability condition)
glow despite its voltage or $pd$ product, is uniquely defined by the underlying topology 
of its secondary ionization.

\section {Discussion}
A key question raised by the foregoing is whether fractal dimension in 
the discharge ionization cascades has any additional meaning besides a 
restatement of the similarity principles and the Townsend breakdown criterion.  
The author would argue that it opens a new avenue for investigation by helping 
us understand that the similarities between discharges due to their primary and 
secondary fractal dimensions helps us to more clearly understand or interpret the 
underlying ionization processes in discharges that would seemingly have no similarity, 
even using the standard similarity principles. The fractal dimensions discussed can be
easily calculated given experimental data for the needed variables. Future suggested research would include 
experimental measurements of both fractal dimensions under different conditions to 
ascertain whether phenomena are similar to or dissimilar from those postulated in this paper. 
One key point the author would like to raise is the scaling that leads to the fractal dimensions 
mentioned in this paper is not related to the scaling in avalanche size in critical phenomena. Since it deals with theoretically infinite avalanches it does not answer nor ask the same questions of avalanche phenomena as theories such as self-organized criticality. It would be interesting, however, to see these techniques applied to a wider class of critical phenomena.
	I would like to thank Robbie Stewart and Steve Hansen for help and suggestions
regarding experimental apparatus and Russell Lyons and Yuval Peres for helpful 
comments on the mathematics of branching sets.

\end{document}